\documentclass[conference]{IEEEtran}

\usepackage{setspace}
\usepackage{amsmath}
\usepackage{graphicx}
\usepackage[linesnumbered,ruled,vlined]{algorithm2e}
\usepackage{multirow}
\usepackage{cite}

\usepackage{lipsum}
\usepackage{fancyhdr}
\begin{document}

\title{Enhanced Estimation of Autoregressive Wind Power Prediction Model Using Constriction Factor Particle Swarm Optimization}

\author{\IEEEauthorblockN{Adnan Anwar, Abdun Naser Mahmood, \textit{Member, IEEE}}
\IEEEauthorblockA{School of Engineering and Information Technology\\
University of New South Wales\\
Canberra, Australia \\
Email: Adnan.Anwar@adfa.edu.au, Abdun.Mahmood@unsw.edu.au}}

\maketitle

\begin{abstract}

Accurate forecasting is important for cost-effective and efficient monitoring and control of the renewable energy based power generation. Wind based power is one of the most difficult energy to predict accurately, due to the widely varying and unpredictable nature of wind energy. Although Autoregressive (AR) techniques have been widely used to create wind power models, they have shown limited accuracy in forecasting, as well as difficulty in determining the correct parameters for an optimized AR model. In this paper, Constriction Factor Particle Swarm Optimization (CF-PSO) is employed to optimally determine the parameters of an Autoregressive (AR) model for accurate prediction of the wind power output behaviour. Appropriate lag order of the proposed model is selected based on Akaike information criterion. The performance of the proposed PSO based AR model is compared with four well-established approaches; Forward-backward approach, Geometric lattice approach, Least-squares approach and Yule-Walker approach, that are widely used for error minimization of the AR model. To validate the proposed approach, real-life wind power data of \textit{Capital Wind Farm} was obtained from Australian Energy Market Operator. Experimental evaluation based on a number of different datasets demonstrate that the performance of the AR model is significantly improved compared with benchmark methods.

\end{abstract}
\IEEEpeerreviewmaketitle

\section{Introduction}

In recent years, renewable energy has gained much popularity and attention because of it's potential in economic and environmental advantages. Some of the benefits include- high stainability, low carbon emission, reduction of environmental impact, saving fuel cost and so on. Other advantages include economical benefits to remote communities and supporting the microgrids during the operation in islanded mode. Although renewable energy, e.g., wind and solar, offers huge benefits~\cite{Anwar2}, their practical use is limited due to their intermittent nature which makes it very challenging to ensure a steady power supply in the grid~\cite{Eriksen1524622, Anwar1}. Because of the variable nature of the renewable energy based power generation sources, transmission and distribution system operators need advanced monitoring and control.

Wind power generation relies on wind speed which varies depending on location and time. For economic and stable operation of the wind power plant, accurate forecasting of wind power is critical. There are two main wind power forecasting approaches, physical method and statistical method. In the first approach, the physical system and power translation processes are modelled in detail. Therefore, physical approaches not only need the information of historical wind speed data but also other information, i.e., meteorological output, hub height of the turbine and physical modelling of power conversion process from wind speed are essential~\cite{Cao4523836}. On the other hand, in a statistical approach, wind power output is modelled as a time-series where the power output at any time instant depends on its previous observation values. The physical approach provides good accuracy for long term forecasting but not so good for short term forecasting as it is computationally very demanding. On the contrary, statistical approaches are well suited for short therm forecasting.

For short term wind power forecasting, different approaches are well studied~\cite{Lei2009915,Hill5966365,Erdem20111405,Poggi20033177,Daniel19911,Venayagamoorthy6269206,Senjyu982201, Methaprayoon4385004}. In a conventional statistical approach, wind power output behaviour is modelled as a time-series. Autoregressive (AR) model has been used for wind energy forecasting in~\cite{Poggi20033177,Daniel19911} and Autoregressive Moving Average (ARMA) model has been used in~\cite{Hill5966365,Erdem20111405}. The Artificial Neural Network (ANN) is also widely used~\cite{Venayagamoorthy6269206,Senjyu982201, Methaprayoon4385004}. However, the ANN based approaches has very slow convergence during the training phase~\cite{ChaoMing1425612}. On the other hand, statistical regressive models are computationally very efficient and widely used for short term forecasting~\cite{Hill5966365,Erdem20111405,Poggi20033177,Daniel19911}.

In the statistical approaches, the forecasting accuracy is highly dependent on the estimated model of the wind power output behaviour. Therefore, it is important to identify the estimated model parameters accurately. Different methods are widely used to estimate the AR model parameters, such as, ordinary Least Squares (LS) approach, Forward Backward (FB) approach, Geometric Lattice (GL) approach and Yule-Walker (YW) approach, etc~\cite{ARmodel,GLAR2,box2008time}. As the wind power output has variable characteristics, the error function obtained from the estimated model may have many local minima. For short-term load forecasting, it has been shown that the Particle Swarm Optimization (PSO), one of the major paradigms of the computational swarm intelligence, converges to the global optimal solution of a complex error surface and finds better solution compared with gradient search based stochastic time-series techniques~\cite{ChaoMing1425612}. Previously, PSO has been widely used in different applications of power system~\cite{Anwar2012,Valle1}. In this work, a modified variant of PSO based on Constriction Factor (CF) is employed to identify the AR parameters more accurately. The proposed CF-PSO based identified AR parameters have better error minimization profiles compared to the well-established LS, FB, GL and YW based approaches.

The organization of this paper is as follows- The formulation of basic PSO and CF-PSO is discussed in Section~\ref{psos}. Autoregressive model order selection and parameter estimation methodology is described in Section~\ref{ARmodel}. The proposed AR Parameter Estimation method based on CF-PSO is illustrated in Section~\ref{PSOmodel}. In Section~\ref{RnD}, results obtained from this experiment are given and compared with four standard techniques. Finally, the paper concludes with some brief remarks in Section~\ref{SecEnd}.

\section{PSO Formulation: Parameters and Variants}\label{psos}
PSO is a multi-objective optimization technique which finds the global optimum solution by searching iteratively in a large space of candidate solutions. The description of basic PSO and CF-PSO formulation is discussed in the following subsections:

\subsection{Basic PSO}

This meta-heuristic is initialized by generating random population which is referred as a swarm. The dimension of the swarm depends on the problem size.
In a swarm, each individual possible solution is represented as a `particle'. At each iteration, positions and velocities of particles are updated depending on their individual and collective behavior.
Generally, objective functions are formulated for solving minimization problems; however, the duality principle can be used to search the maximum value of the
objective function~\cite{deb2004optimization}.
At the first step of the optimization process, an \emph{n}-dimensional initial population (swarm) and control parameters are initialized. Each particle of a swarm is
associated with the position vector and the velocity vector, which can be written as\\

velocity vector, $\textbf{v}_{i} = [{v}_{i}^{1},{v}_{i}^{2},...,{v}_{i}^{n}]$
and

position vector, $\textbf{x}_{i} = [{x}_{i}^{1},{x}_{i}^{2},...,{x}_{i}^{n}]$     \\

where n represents the search space dimension. Before going to the basic PSO loop, the position and velocity of each particle is initialized. Generally, the initial position of the ${i}^{th}$ particle
${x}_{i}$ can be obtained from uniformly distributed random vector U (${x}_{min}, {x}_{max}$), where ${x}_{min}$ and ${x}_{max}$ represents the lower and upper limits of the solution space respectively.

During the optimization procedure, position of each particle is updated using~(\ref{peq})
\begin{equation}
\textbf{x}_{i}^{t+1} = \textbf{x}_{i}^t+\textbf{v}_{i}^{t+1}
\label{peq}
\end{equation}
where ${x}_{i}\in{R}^{n}$ and ${v}_{i}\in{R}^{n}$.
\\

At each iteration, new velocity for each particle is updated which drives the optimization process. The new velocity of any particle is calculated based on
its previous velocity, the particle's best known position and the swarm's best known position. Particle's best known position is it's
location at which the best fitness value so far has been achieved by itself and swarm's best known position is the location at which the best fitness value
so far has been achieved by any particle of the entire swarm~\cite{EberhartConsPSO}. The velocity equation drives the optimization process which is updated using~(\ref{veq})

\begin{equation}                                                                                                                                                                                    \textbf{v}_{i}^{t+1} = w.\textbf{v}_{i}^t+ {r}_{1}.{c}_{1}.(\textbf{p}_{i}-\textbf{x}_{i}^t)+ {r}_{2}.{c}_{2}.(\textbf{p}_{g}-\textbf{x}_{i}^t)
\label{veq}
\end{equation}

In this equation, \emph{w} is the inertia weight. $(\textbf{p}_{i}-\textbf{x}_{i}^t)$ represents the
`self influence' of each particle which quantifies the performance of each particle with it's previous performances. The component $(\textbf{p}_{g}-\textbf{x}_{i}^t)$ represents
the `social cognition' among different particles within a swarm and quantify the performance relative to other neighboring particles.
The learning co-efficients ${c}_{1}$ and ${c}_{2}$ represent the trade-off between the self influence part and the social cognition part of the particles~\cite{Zeineldin2011}.
The values of ${c}_{1}$ and ${c}_{2}$ are adopted from previous research and is typically set to 2~\cite{Etemadi1}.
In eqn~(\ref{veq}), ${P}_{i}$ is particle's best known position and ${P}_{g}$ is swarm's best known position.

In the solution loop of PSO, the algorithm continues to run iteratively, until one of the following stopping conditions is satisfied~\cite{engelbrecht2007computational}.
\begin{enumerate}
\item Number of iterations reach the maximum limit, e.g., 100 iterations.
\item No improvement is observed over a number of iterations, e.g., error less than $\epsilon$~=~0.001
\end{enumerate}

\subsection{Constriction factor PSO (CF-PSO) with boundary conditions}

To achieve better stability and convergent behavior of PSO, a constriction factor has been introduced by Clerc and Kennedy in~\cite{Clerc2002}. The superiority of CF-PSO over inertia-weight
PSO is discussed in~\cite{EberhartConsPSO}. Basically, the search procedure of CF-PSO is improved using the eigenvalue analysis and the system behavior can be controlled which ensures a
convergent and efficient search procedure~\cite{Vlachogiannis2006}. To formulate CF-PSO, (\ref{veq}) is replaced by~(\ref{veqcons})-(\ref{kcons2})\cite{Valle1,Renato2006,Heo2006,Naka2003}.

\begin{equation}
\textbf{v}_{i}^{t+1} = \emph{k}~[\textbf{v}_{i}^t+ {r}_{1}.{c}_{1}.(\textbf{p}_{i}-\textbf{x}_{i}^t)+ {r}_{2}.{c}_{2}.(\textbf{p}_{g}-\textbf{x}_{i}^t)]
\label{veqcons}
\end{equation}

where

\begin{equation}
\emph{k}= \frac{2}{| ~2- \varphi - \sqrt {\varphi^2-4\varphi} ~|}
\label{kcons}
\end{equation}

and

\begin{equation}
\varphi=  {c}_{1} +{c}_{2},~ ~   \varphi > 4
\label{kcons2}
\end{equation}

here the value of $\varphi$ must be greater than 4 to ensure a stable and convergent behavior~\cite{EberhartConsPSO,Clerc2002}. Usually, the value of $\varphi$ is set to 4.1 (${c}_{1} ={c}_{2}=2.05$); therefore, the value of \emph{k} becomes 0.7298~\cite{Renato2006,Heo2006,Naka2003}.

Boundary condition~\cite{Shenheng2007}, which helps to keep the particles within allowable solution space, is also applied in this research as shown below:
\begin{equation}
{x}_{i}^{t}= \left\{ \begin{array}{ll}
 ~~{x}_{i}^{max} & \textrm{if ${x}_{i}^{t}>{x}_{i}^{max}$}\\
 ~~{x}_{i}^{min} & \textrm{if ${x}_{i}^{t}<{x}_{i}^{min}$}\
\end{array} \right.
\label{vclp}
\end{equation}\\

\section{Autoregressive (AR) Model}\label{ARmodel}

AR is a univariate time-series analysis model that is widely used for model estimation and forecasting. In an AR model, the output variable has linear association with its own previous observations.
For a sample period of $t=[1,2,...,t]^T$, a $\rho$-order AR model can be written following the expression below~\cite{Poggi20033177,box2008time}:
\begin{equation}\label{AReq}
  X_t = C + \sum_{i=1}^{\rho} \varphi_i X_{t-i}+ \varepsilon_t
\end{equation}
where, $\varphi_1, \ldots, \varphi_p$ are the lag parameters of the model, $C$ is the constant term, and $\varepsilon_t$ is the white gaussian noise with zero mean.

To select the minimal appropriate lag order $\rho$, Akaike information criteria is used following~(\ref{AICeq}), where $M$ is the number of parameters in the AR model, $n$ is the effective number of observations, ${\hat{\sigma_a}}$ is the maximum likelihood of the estimate of the error covariance~\cite{Peiyuan5340622,wei2006time}. The best fitted model has the minimum AIC value.

\begin{equation}\label{AICeq}
  AIC(\rho)=n~ln({\hat{\sigma_a}}) +2{M}
\end{equation}\\

\section{Estimation of the Autoregressive Model Parameters using CF-PSO}\label{PSOmodel}

Firstly, the structure of the AR model is selected based on~(\ref{AReq}) for a predetermined order $\rho$. After that the model parameters are estimated using CF-PSO. During the optimization loop, the algorithm determines the optimal parameters by minimizing the Residual Sum of Squares (RSS) of the estimated model as shown in (\ref{RSSeq}), where $AD$ is the actual data that need to be predicted and $ED$ is the estimated data.
\begin{equation}\label{RSSeq}
RSS = \sum_{i=1}^n (AD_i - ED_i)^2
\end{equation}
The steps of CF-PSO based AR Parameter Estimation are discussed below.
\begin{enumerate}

\item Initialize particle's position $\textbf{x}_{i}$\ and velocity $\textbf{v}_{i}$ in an \emph{n}-dimensional search space. Here, the dimension \emph{n} represents the order $\rho$ of the AR model and each vector of the particle position indicates a potential solution.

\item Calculate the RSS for each `initial' particle positions and determine the particle's best known position $\textbf{P}_{i}$ and swarm's best known position $\textbf{P}_{g}$.

\item Update particle's position $\textbf{x}_{i}$ and velocity $\textbf{v}_{i}$ following (\ref{peq}) and (\ref{veq}) if necessary.

\item Calculate the RSS with the updated velocity $\textbf{v}_{i}$ and position $\textbf{x}_{i}$.

\item Repeat (3) and (4) until the stopping criteria is satisfied, i.e., there is not significant change in RSS over a good number of iterations or the algorithm reaches it's maximum iteration limit. Here maximum iteration is considered 100.

\end{enumerate}

\begin{figure}
    \centering
    \includegraphics[width=3.2in]{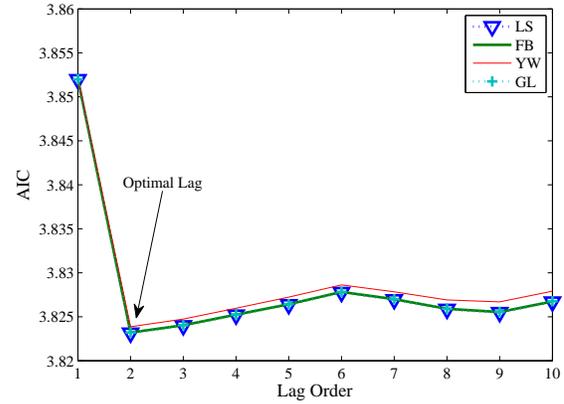}
    \caption{Optimal LAG selection}
    \label{AIClbl}
\end{figure}

\section{Results and Discussion}\label{RnD}

The proposed CF-PSO based AR model parameter estimation algorithm is implemented on the practical wind power output data of the \textit{Capital Wind Farm}, obtained from the Australian Energy Market Operator~\cite{AEMOonline}.
The algorithm is implemented using Matlab and standard LS, FB, GL and YW based approaches are evaluated using `System Identification Toolbox'~\cite{Mident}.

To compare the performance of the proposed method and four aforementioned well-established methods, following evaluation indices are used:
\begin{enumerate}
\item Mean Square Error (MSE):
\begin{equation}
  MSE=\frac{1}{n}\sum_{i=1}^n(\hat{Y_i} - Y_i)^2
\end{equation}
where $Y_i$ is the actual data and $\hat{Y_i}$ is the data from the estimated model.

\item Akaike's Final Prediction Error (FPE)~\cite{ChaoMing1425612}:
\begin{equation}
  FPE= \left(\frac{1+h/H}{1-h/H}\right)\times\nu
\end{equation}
where, $h$ is the number of parameters in the estimated model, $H$ is the number of values of the model and $\nu$ is the loss function.

\item  Normalized Mean Square Error (NMSE)~\cite{Tay2002847}:
\begin{equation}
\begin{split}
  NMSE=\frac{1}{\delta^2n} \sum_{i=1}^n[{Y_i} - \hat{Y_i}]^2\\
   {\delta^2}=\frac{1}{n-1} \sum_{i=1}^n[{Y_i} - mean({Y_i})]^2\\
  \end{split}
\end{equation}
where $Y_i$ is the actual data and $\hat{Y_i}$ is the data from the estimated model. The value of NMSE varies between `-inf' to `1'. While `-inf' indicates a bad fit, 1 represents the perfect fit of the data.

\end{enumerate}

Firstly, the appropriate lag order is determined. In order to do that the value of ${\rho}$ is varied from $0$ to ${\rho}_{max}$ and for each value of ${\rho}$, the AIC is calculated following the information criterion in~(\ref{AICeq}). Once all AIC value is known, ${\rho}$  is selected for that fitted AR(${\rho}$) model which leads to the minimum AIC value. Considering ${\rho}_{max}=10$, in this analysis AIC value is observed when ${\rho}=2$ for all four standard methods as shown in Fig.~\ref{AIClbl}. Now, the wind power output data (first week of March 2012 with 5 minute interval) of the `capital Wind Farm' is used to evaluate the performance of the five approaches including the CF-PSO based AR model. For the proposed method, the results are documented considering the mean value of 30 individual runs. The obtained results are summarized in the Table.~\ref{case1results} and the best results among all approaches are highlighted. Results presented in Table.~\ref{case1results} shows that the proposed method outperforms the other approaches for this test case.
Considering the error value (MSE and FPE) of the LS approach as a base scenario, the Error Minimization Performance (EMP) of other approaches are evaluated following~(\ref{myindex}), where $Error_{LS}$ is the MSE or FPE of the LS method and $Error_i$ is the corresponding MSE or FPE of other approaches. The positive value of EMP indicates an improvement of error minimization performance
over LS approach while negative value represents that the performance is worse than the LS approach.

\begin{equation}\label{myindex}
EMP= \frac{ Error_{LS} - Error_i }{Error_{LS}}\times 100 \%
\end{equation}

In order to justify the performance, the proposed method is employed considering another data set, second week 5s interval data of the month March 2012 for capital wind farm. Results from this method is shown in Table~\ref{case1results2}. In this test case, the proposed CF-PSO based AR model reduces the error indices most. Compared with LS, almost 40\% reduction of error is achieved for this test data. Moreover, the NMSE for the standard LS, FB, GL and YW based approaches are around 96.7\% while proposed method experiences 98.9\%, as shown in Table.~\ref{case1results2}. As the value of NMSE close to unity indicates the best performance, proposed method also shows it superiority for this test case.
\begin{table}
\caption{Performance of AR Parameter Estimation Considering The First Week data of March 2012}\label{case1results}
\footnotesize
\centering
\begin{center}
\begin{tabular}{| c | c | c| c| c| c|}
\hline
Method	&	MSE	&	$EMP_{MSE}$	&	FPE	&	$EMP_{FPE}$	&	NMSE	\\
&&(\%)&&(\%)&\\ \hline
LS	&	46.7133	&	$-$	&	46.806	&	$-$	&	0.975	\\ \hline
FB	&	46.7153	&	-0.004	&	46.808	&	-0.004	&	0.9751	\\ \hline

YW	&	46.7441	&	-0.066	&	46.837	&	-0.066	&	0.9751	\\ \hline
GL	&	46.7153	&	-0.004	&	46.808	&	-0.004	&	0.9751	\\ \hline
CF-PSO	&	\textbf{45.612}	&	\textbf{+2.357}	&	\textbf{45.703}	&	\textbf{+2.357}&	\textbf{0.9757}	\\ \hline
\end{tabular}
\end{center}
\end{table}

 \begin{table}
\caption{Performance of AR Parameter Estimation Considering The Second Week data of March 2012}\label{case1results2}
\footnotesize
\centering
\begin{center}
\begin{tabular}{| c | c | c| c| c| c|}
\hline
Method	&	MSE	&	$EMP_{MSE}$	&	FPE	&	$EMP_{FPE}$	&	NMSE	\\
&&(\%)&&(\%)&\\ \hline
LS	&	17.17	&	$-$	   &	17.206	&	$-$	   &	0.9669	\\ \hline
FB	&	17.18	&	-0.0623	&	17.217	&	-0.062	&	0.9669	\\ \hline

YW	&	17.14	&	0.182	&	17.175	&	0.182	&	0.9669	\\ \hline
GL	&	17.18	&	-0.067	&	17.218	&	-0.067	&	0.9669	\\ \hline
CF-PSO	&	\textbf{10.22}	&	\textbf{+40.487}	&	\textbf{10.240}	&	\textbf{+40.487}	&	\textbf{0.9889}	\\ \hline

\end{tabular}
\end{center}
\end{table}

\begin{figure*}
    \centering
    \includegraphics[width=6.2in,height=3in]{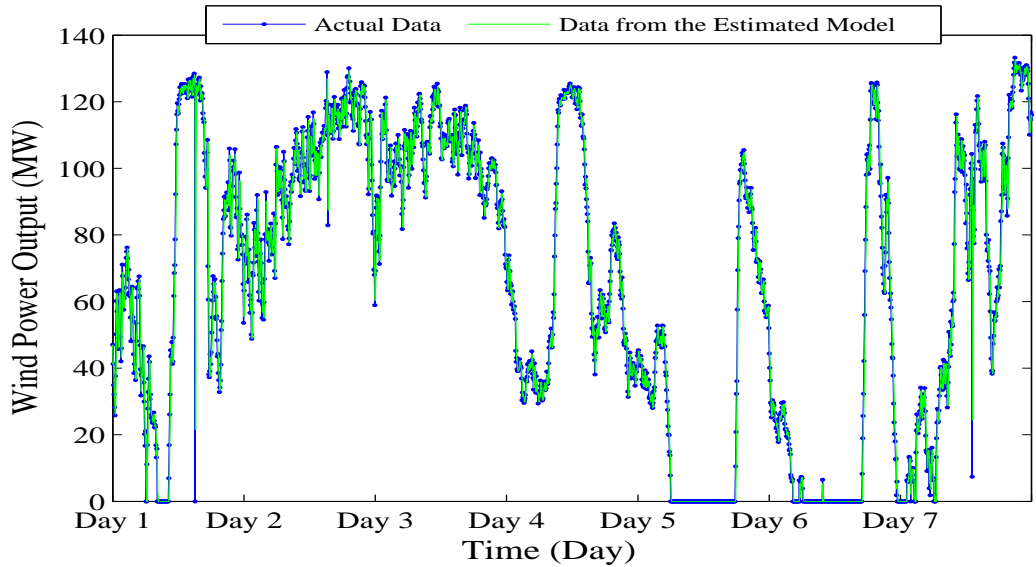}
    \caption{Comparision of actual data with the estimated model data}
    \label{actest}
\end{figure*}

\begin{figure}[h]
    \centering
    \includegraphics[width=3.2in]{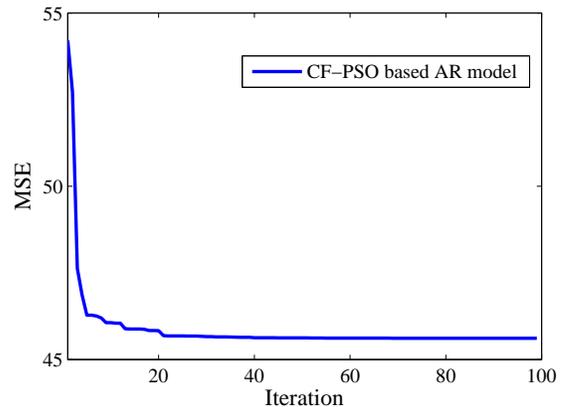}
    \caption{The convergence characteristics of the CF-PSO based proposed algorithm}
    \label{cfpsolbl}
\end{figure}

Fig.~\ref{actest} shows the actual wind output data and the estimated model data using the proposed method. The convergence characteristics of the CF-PSO based proposed algorithm is shown in Fig.~\ref{cfpsolbl}. From the figure, the algorithm convergence within 40 iterations.
According to the results shown in Table.~\ref{case1results} and Table.~\ref{case1results2}, the best improvement is observed considering the CF-PSO based AR modelling among these five approaches for both of the test data sets. The CF-PSO based AR parameter estimation method finds a better solution compared to the gradient based methods due to its global search capabilities. It is important to mention that these well-established
gradient based methods may trap in local minima as referred by Huang et al~\cite{ChaoMing1425612}. On the other hand, CF-PSO
finds a global optimal solution. In our analysis, we found that the performance of the CF-PSO varied based on the wind data characteristics. If
wind data has a global minima that is very close to the local minima, the performance of the CF-PSO is slightly
improved compared with other algorithms (as observed in Table. I). On the other hand, if the local minimum is
far from the global minima value, significant improvement is observed using the CF-PSO algorithm (as
observed in Table. II).

\section{Conclusion}\label{SecEnd}

In this paper, constriction factor based PSO is employed to enhance the performance of the time-series autoregressive model. The proposed algorithm is implemented to estimate the wind power output considering practical wind data. Using the global search capable CF-PSO based proposed AR model, the results obtained in this experiment show that the algorithm finds the solution very accurately and efficiently (within 40 iterations). To justify the results obtained from the proposed method, four algorithms including the widely used Least-Square method and Yule-Walker method are employed for comparison. Experimental results conducted in this experiments show that the proposed method enhances the AR estimation model with better accuracy compared to other four well-established method. In this experiment, the exogenous input variables are not considered during the model estimation, which will be included in the future work. Since the proposed model enhances the performance of the autoregressive model by minimizing model estimation errors more effectively, in the future work the forecasting performance will also be explored in detail.


{}



\end{document}